\documentclass[aps,prx,notitlepage,superscriptaddress,longbibliography,twocolumn,nofootinbib,floatfix]{revtex4-2} 

\usepackage{Packages}

\begin{document}
\title{Su-Schrieffer-Heeger-Hubbard model at quarter filling:\\
effects of magnetic field and non-local interactions}

\author{David Mikhail}
\affiliation{School of Physics, University of Melbourne, Parkville, VIC 3010, Australia}
\author{Stephan Rachel}
\affiliation{School of Physics, University of Melbourne, Parkville, VIC 3010, Australia}

\date{\today}

\begin{abstract}
The interplay and competition of topology and electron\nd electron interactions have fascinated researchers since the discovery of topological insulators. The Su\nd Schrieffer\nd Heeger\nd Hubbard (SSSHH) model is a prototypical model which includes both non-trivial topology and interactions. Due to its simplicity, there are several artificial quantum systems which can realize such a model to a good approximation. 
Here we focus on the quarter-filled case, where interactions and dimerization open a charge gap. In particular, we study the single-particle spectral function for the extended SSHH model with magnetic field and explore several parameter limits where effective model descriptions arise. In the strongly-dimerized limit, we show that the low-energy excitations of the spectral function resemble a half-filled Hubbard model with effective dimer sites and renormalized couplings. 
For strong magnetic field and interactions, we find physics akin to the spinless Su\nd Schrieffer\nd Heeger model at half filling, featuring a non-interacting topological phase transition. Moreover, in light of the recent realization of this model in quantum dot simulation, we provide evidence for the stability of the topological phase towards moderate non-local interactions in the experimentally expected parameter range.      
\end{abstract}

\maketitle

\section{Introduction}
Topological phases of matter have been highly sought after since the discovery of the integer quantum Hall effect\,\cite{Klitzing1980a}, due to their unique properties and potential applications in quantum technologies. 
Topological insulators (TI) \md characterized by topologically non-trivial band structure with a bulk gap and conducting boundary states\md in particular, have attracted substantial interest as they do not require strong magnetic fields\,\cite{Haldane1988}. 
The theoretical prediction\,\cite{Kane2005b,Kane2005a,Bernevig2006b,Bernevig2006} and subsequent discovery\,\cite{Koenig2007} of $\mathbb{Z}_2$-TIs, which are protected by time-reversal symmetry, sparked efforts to classify non-interacting Hamiltonians into topological equivalence classes called the periodic table of topological insulators and superconductors\,\cite{Schnyder2008a,Kitaev2009bc}. 
    
However, many real-world materials exhibit non-negligible interactions that cannot be captured by such single-particle descriptions.
While the effects of interactions on topological phases have been the subject of extensive research\,\cite{Hohenadler2013,Rachel2018a}, a comprehensive understanding remains elusive.
This gap highlights the importance of studying strongly correlated systems to fully grasp the interplay between topology and interactions.
    
Analogue quantum simulation offers a promising pathway to explore these systems in controlled environments\,\cite{Esslinger2010,Georgescu2014a}, paving the way for deeper insights and technological advancements.
Topological phases have been investigated across various platforms, including classical RLC circuits\,\cite{Lee2018}, cold atoms in optical lattices\,\cite{Atala2013,Aidelsburger2013,Jotzu2014a,Meier2016,Cooper2019a,Sompet2022}, superconducting circuits\,\cite{Flurin2017,Cai2019,Tan2021}, trapped ions\,\cite{Bermudez2011,Schmied2011,Manovitz2020}, and semiconductor quantum dot arrays\,\cite{Le2020c,Kiczynski2022bc}. 
The latter, in particular, enable the study of strong electron-electron interactions ($U/t\sim 4\nd14$) at low temperatures ($T\sim t/420$)\,\cite{Salfi2016}, revealing rich fermionic many-body quantum phenomena\,\cite{Salfi2016,Hensgens2017o,Le2017,Dusko2018c,Wang2021}. 
    
Recently, a one-dimensional (1D) interacting topological phase was simulated on a 10-site chain of phosphorus donors in silicon (Si:P)\,\cite{Kiczynski2022bc}, which are known to exhibit strong Coulomb confinement ($U\sim 25 \,\rm meV$).
The precise placement of dopants on the sub-nanoscale using a scanning tunneling microscope (STM), a process called STM-lithography\,\cite{Shen1995,Schofield2003,Fuechsle2012b}, allows for high control over the hopping amplitudes $t_{ij}\sim 1\nd 5 \,\rm meV$\,\cite{Kiczynski2022bc} between neighboring quantum dots at positions $\vec R_i$ and $\vec R_j$, respectively, setting the energy scale for effective temperature $T/t$ and on-site interactions $U/t$\,\cite{Salfi2016}. 
In this experiment\,\cite{Kiczynski2022bc}, the interacting Su-Schrieffer-Heeger (SSH) model\,\cite{Su1979b}\md a canonical model of a one-dimensional topological insulator\md was simulated. The study successfully probed zero-energy edge states, associated with the topological phase, through conductance measurements.
    
The SSH model\,\cite{Su1979b}, originally developed to describe the electronic properties of polyacetylene, provides a simple yet powerful framework for studying topological insulators. 
Its tight-binding Hamiltonian models spinful electrons on a dimerized chain. The model captures the essence of topological insulators, the bulk-boundary correspondence, by exhibiting topologically protected edge states at zero energy, which arise due to the presence of a non-trivial topological invariant in its band structure. 
    
In dopant arrays, STM methods can be employed to measure local transport properties\,\cite{Salfi2016,Voisin2020}. 
In particular, scanning tunneling spectroscopy (STS) can directly probe spatial and spectral properties of edge excitations by measuring the local density of states (LDOS)\,\cite{Cheon2015,Drost2017,Schneider}. Here we focus on the single-particle spectral function (SPSF) in real space, which generalizes the LDOS to interacting systems\,\cite{Secchi2012c,Ervasti2017c}.    
    
By means of exact diagonalization (ED), we calculate the SPSF of the extended Su-Schrieffer-Heeger-Hubbard (SSHH) model\,\cite{Manmana2012,Yoshida2014,Wang2015f,Ye2016,Le2020c,Mikhail2022d}. As a model Hamiltonian for Bechgaard salts, it is also referred to as the (extended) Peierls-Hubbard model in the literature\,\cite{Penc1994,Favand1996,Gebhard1997c,Nishimoto2000,Nishimoto2001,Mila2001,Tsuchiizu2001,Otsuka,Rissler2005,Dzierzawa2005a,Grage2005a,Ejima2007a,Ejima2016}.
    
While the half-filled case of the SSHH model has been studied extensively, see Sec.\,\ref{sec:model} for a brief discussion, the recent experiment claims to operate at quarter filling\,\cite{Kiczynski2022bc}. 
Thus we will focus in this paper on the rarely studied quarter-filled case.

Most STMs are equipped with magnetic fields up to a few Tesla. This might be an important experimental knob to substantiate the existence of a topological phase and probe correlation physics. 
We take this as an additional motivation to investigate the effect of a magnetic field on the SPSF in the SSHH model. 
We observe the previously predicted transition\,\cite{Le2020c} between the topologically trivial SSHH model and the topological spinless (non-interacting) SSH model driven by a strong Zeeman field. 

Last but not least, in the recent experiment\,\cite{Kiczynski2022bc}, it was further suggested that the involved electron\nd electron interactions are not purely local; small but possibly non-negligible nearest-neighbor contributions might be present. 
We will also study the effect of additional nearest-neighbor repulsion on the SPSF of the SSHH model at quarter filling.
Moreover, we find that the observed spin-polarized SSH model is robust against moderate nearest-neighbor interactions at a scale relevant for experiments.  

The rest of this paper is organized as follows. The observables, such as spectral function and charge correlation functions, are introduced in Sec.\,\ref{sec:method}. In Sec.\,\ref{sec:model} we introduce the different components of the SSHH model and briefly discuss its properties at half filling (\ref{sec:hf}), before we address the quarter-filled case in Sec.\ref{sec:qf}. 
In Sec.\,\ref{sec:ub} we present results for the SSHH model in the presence of a magnetic field, followed by calculations for the extended SSHH model in Sec.\,\ref{sec:uv}. 
The combined effect of nearest-neighbor interactions and magnetic field is analysed in Sec.\,\ref{sec:ubv}. Finally, our main results are summarized in Sec.\,\ref{sec:conclusion}.

\section{Method}\label{sec:method}
All numerical observables in this sections are calculated using Lanczos-based ED for a system of $L=12$ sites. 
In the weak-tunneling regime the differential conductance $dI/dV$ in STS measurements is well approximated by the single-particle charge excitation spectrum\,\cite{Ervasti2017c}. The SPSF is the probability distribution of these charge excitations and, in the lattice site basis, given by 
\begin{align}
    A_{i,\sigma}(\omega)= A^-_{i,\sigma}(\omega) + A^+_{i,\sigma}(\omega)    
    \label{spsf}.
\end{align}
where $A^{\pm}_{i,\sigma}(\omega)$ describes the addition and removal of an electron, respectively, on site $i$ and at energy $\omega$.  
At zero temperature, the removal and addition spectral functions can be written as 
\begin{align}
    A^-_{i,\sigma}(\omega) &= \sum_n |\braket{\Psi^{N-1}_n|\hat{c}_{i,\sigma}|\Psi^{N}_0}|^2 \delta\left(\omega - \left( E^N_0 - E^{N-1}_n\right)\right), \label{sfm}\\
    A^+_{i,\sigma}(\omega) &= \sum_n |\braket{\Psi^{N+1}_n|\hat{c}^\dagger_{i,\sigma}|\Psi^{N}_0}|^2 \delta\left(\omega - \left( E^{N+1}_n - E^N_0 \right)\right),\label{sfp}
\end{align}
where $\ket{\Psi^N_n}$ denotes the $n$th many-body eigenstate occupied by $N$ electrons and with eigenergy $E_n^N$.  
Using the identity for $\delta$-functions $\delta(x) = - \rm Im \left[ \lim_{\eta \to 0+}1/(x+i\eta)\right]/\pi$, it can be shown\,\cite{Gagliano} that Eqs.\,\eqref{sfm} and~\eqref{sfp} are equivalent to  
\begin{align}
    A^-_{i,\sigma}(\omega) &=  -\lim_{\eta \to 0+}\frac{1}{\pi}\rm Im\braket{\hat{c}^\dagger_{i,\sigma}\frac{1}{z^{-} - \hat{H}} \hat{c}_{i,\sigma} } \label{sfm2}\\
    A^+_{i,\sigma}(\omega) &= -\lim_{\eta \to 0+}\frac{1}{\pi}\rm Im\braket{ \hat{c}_{i,\sigma}\frac{1}{z^{+} - \hat{H}} \hat{c}^\dagger_{i,\sigma} } \label{sfp2}
\end{align}
respectively, where $\braket{\dots}\equiv\braket{\Psi^N_0|\dots|\Psi^N_0}$ denotes the expectation value taken in the ground state,  $z^\pm = w \mp E^N_0 + i \eta$ and broadening $0 <\eta \ll 1$. Here we choose $\eta=0.025$. We evaluate the spectral functions Eqs.\,\eqref{sfm2} and~\eqref{sfp2} as continued fractions using the Lanczos procedure\,\cite{Gagliano,Fulde1991}.

\begin{figure*}[t!]
\begin{center}
    \centering
    \vspace*{0.5cm}   
    \includegraphics[width=\linewidth]{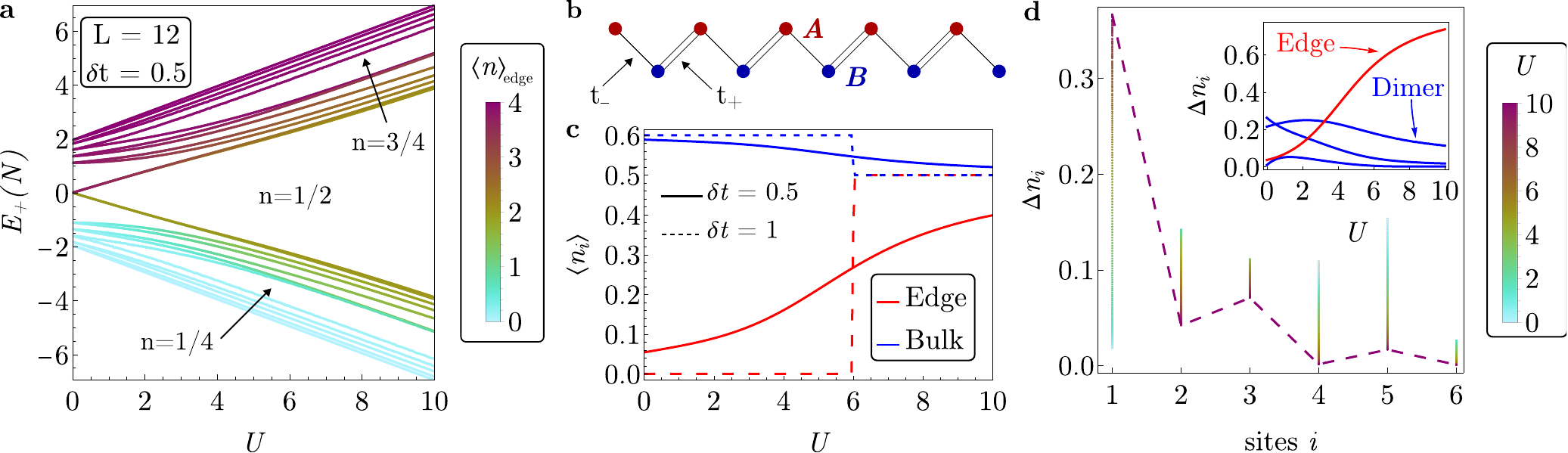}
    \caption{\textbf{Addition spectrum and charge density for a chain of $L=12$ sites and dimerization $\delta t = 0.5$.} \textbf{a} Addition energies $E_+(N)$ in dependence of $U$ for $N=1,\dots, 2L$. The color indicates the occupation of both edges $\braket{\hat{n}}_\text{edge} = \braket{\hat{n}_1} + \braket{\hat{n}_L}$.\ \textbf{b} Schematic of a topological SSH chain with hopping amplitudes $t_{\pm}= t \pm \dt $.\ \textbf{c} Average occupation of an edge (red) and bulk (blue) site in dependence of $U$ at quarter filling (here $N=6$). The dashed curves are calculated in the fully dimerized limit $\dt =1$ while solid lines are calculated for $\dt = 0.5$.\ \textbf{d} Local charge density distribution $\Delta n_i = \braket{\hat{n}_i(N=7)}-\braket{\hat{n}_i(N=6)}$ in dependence of $U$. Inset: Charge density distribution for both edges (red) and first three bulk dimers (blue).}\label{f:u}
\end{center}
\end{figure*}

\section{Model}\label{sec:model}
We focus on the SSHH model, which is the interacting variant of the SSH model. The latter is described by the tight-binding Hamiltonian 
\begin{align}
    \hat{H}_{\text{SSH}}= \sum_{i,\sigma=\uparrow,\downarrow} \left(t+(-1)^i \delta t \right) \left[\hat{c}^\dagger_{i+1,\sigma}\hat{c}_{i,\sigma} + \text{H.c.}\right],
    \label{hssh}
\end{align}
with creation operator $\hat{c}^\dagger_{i,\sigma}$ of an electron on site $i$ and spin $\sigma$, nearest-neighbor hopping amplitude $t$ and chain dimerization $\delta t$. Henceforth, we present energies in units of $t\equiv 1$. 

The SSH chain has two possible configurations which are characterized by the position of strong and weak bonds and which depend on the sign of $\dt$. A sketch of an SSH chain for open boundary conditions (OBC) with positive dimerization $\dt>0$ is shown in Fig.\,\ref{f:u}\,b. Strong bonds with hopping $t_{+}=t+\dt$ are double lines while weak bonds with hopping $t_{-}=t-\dt$ are shown as single lines. The depicted configuration (Fig.\,\ref{f:u}\,b) shows the topological regime; the two edge sites are weakly coupled to the bulk and their single-particle eigenstates are exponentially localized in real space. Their overlap with bulk sites vanishes in the thermodynamic limit. 

Due to chiral symmetry, the spectrum of \eq{hssh} is symmetric around $E=0$. For $L\to \infty$, the single-particle edge states become true eigenstates of the Hamiltonian \eq{hssh} with eigenenergies pinned to zero. At finite system sizes, the edge state wavefunctions hybridize and their energies are small but finite.   

The localized edge states are protected by the chiral symmetry which ensures the presence of a bulk gap for finite $\dt$. The topological nature of these states can be inferred from the Berry phase $\phi$ (or winding number $\nu$) which jumps from $\phi = 0 \to \pi$ ($\nu = 0 \to 1$) when $\dt<0 \to \dt>0$. At $\dt=0$ the bulk gap closes, marking the critical point of the topological phase transition.  

To incorporate electron-electron correlations, we extend the SSH model by local and non-local Hubbard terms, \eq{hu} and \eq{hv}, respectively. The magnetic field is modeled by a Zeeman term \eq{hb}, which couples to the $z$-component of the spin. The total many-body Hamiltonian is given by
\begin{align}
    \hat{H} = \hat{H}_{\text{SSH}} + \hat{H}_U + \hat{H}_V + \hat{H}_B,
    \label{htot}
\end{align}
with
\begingroup
\addtolength{\jot}{1em}
\begin{align}   
    \hat{H}_U &= U\sum_i \hat{n}_{i,\uparrow}\hat{n}_{i\downarrow} \label{hu},\\
    \hat{H}_V &= V \sum_{i} \hat{n}_{i}\hat{n}_{i+1} \label{hv},\\
    \hat{H}_B &= - B\sum_i  (\hat{n}_{i,\uparrow} - \hat{n}_{i,\downarrow} ) \label{hb}. 
\end{align}
\endgroup
Here, the fermionic number operator $\hat{n}_i=\sum_\sigma \hat{n}_{i,\sigma}=\sum_\sigma \hat{c}^\dagger_{i,\sigma}\hat{c}_{i,\sigma}$ counts the occupation of site $i$. The coupling constants $U$, $V$ and $B$ refer to on-site interaction, nearest-neighbor interaction and coupling to the magnetic field, respectively. The latter is related to the magnetic field $h$ via $B= g\mu_B h$ with the dimensionless electron g-factor $g\approx2$ and the Bohr magneton $\mu_B = e\hbar/2m$. Since $\mu_B = 1/2$ in atomic units, we can identify $B$ with the magnetic field $h$ for the purpose of our calculations. All couplings in \eq{htot} are always given in units of the hopping $t=1$.

\subsection{SSHH model at half filling}\label{sec:hf}
The SSHH model has been extensively studied in the past.
At half filling, the topological phase transition of the SSH model is known to persist for finite $U>0$, evidenced by a change of the Green's function-derived bulk winding number\,\cite{Volovik2003} upon sign change of the dimerization\,\cite{Manmana2012}. Furthermore, a persistent non-zero degeneracy of the entanglement spectrum\,\cite{Yoshida2014,Ye2016} suggests adiabatic continuity between the $U=0$ and $U>0$ cases, respectively. 

As on-site interactions introduce a Mott gap in the charge sector at half filling, the phase transition is signalled in the bulk (PBC) by a closing and reopening of the many-body gap associated with spin excitations\,\cite{Yoshida2014}. 
For OBC, the single-particle edge states of the SSH model are gapped out by $U$. Instead, the bulk-boundary correspondence is realized by collective gapless spin excitations\,\cite{Manmana2012,Yoshida2014} of local edge spins which are uncorrelated with the bulk\,\cite{Le2020}, but correlated with each other\,\cite{Wang2015f,Barbiero2018c}. 

\begin{figure*}[t!]
\begin{center}
    \centering
    \vspace*{0.5cm}   
    \includegraphics[width=\linewidth]{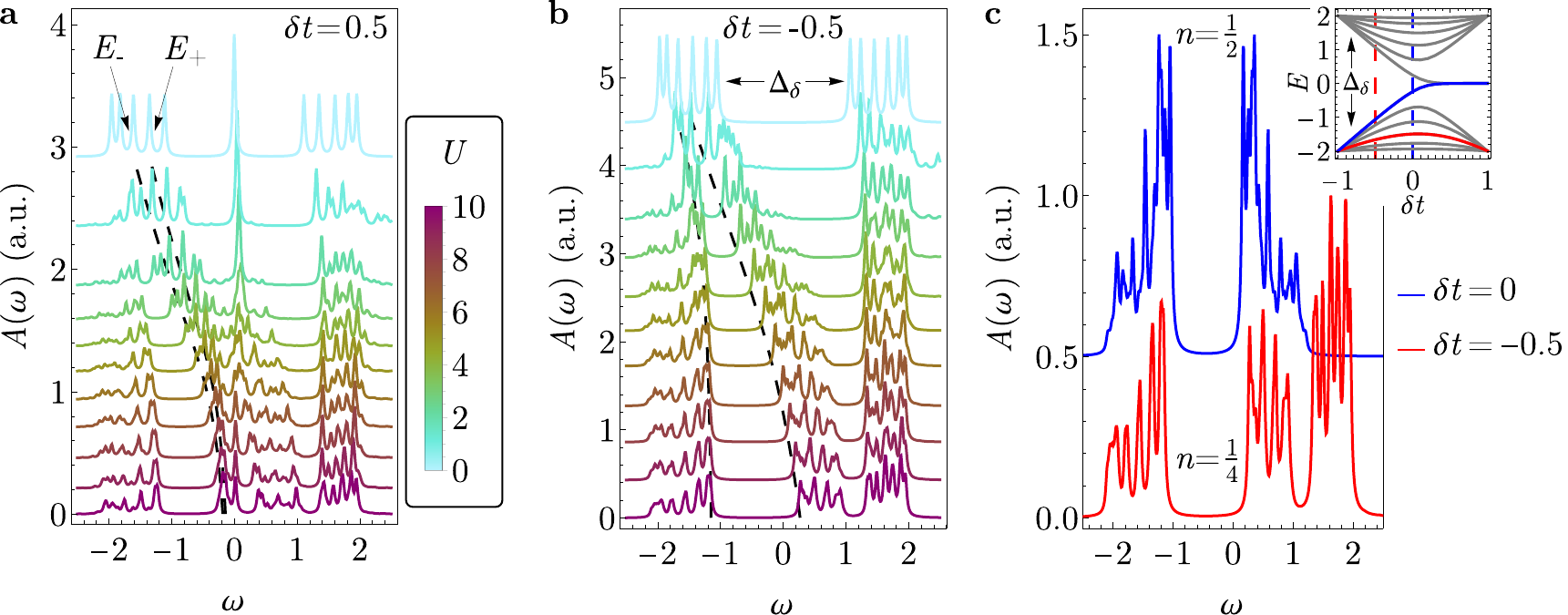}
    \caption{\textbf{Spectral function of quarter filled SSHH model.} \textbf{a} Positive dimerization $\dt=0.5$, with increasing on-site interaction $U$, starting from the top at $U=0$. The addition and removal energies at quarter filling $E_+$ and $E_{-}$ are indicated by dashed black lines.\ \textbf{b} Same as \textbf{a} but for negative dimerization $\dt=-0.5$. Dimerization gap $\Delta_\delta = 4\dt$ is indicated for the non-interacting SSH model.\ \textbf{c} Comparison between density of states of the half-filled (non-dimerized) Hubbard model (blue) with hopping $(t-\dt)/2$ and $U=1.85$ and quarter filled SSHH model (red) at $U=10$ and $\dt=-0.5$. The center of both charge gaps is aligned for visibility. The energy spectrum for the $L=12$ site SSH chain is depicted in the inset in dependence of $\dt$. The red and blue solid lines indicate single-particle energies at quarter and half filling, respectively. The dashed lines mark the dimerization values of the density of states plots. }\label{f:u2}
\end{center}
\end{figure*}
    
The SPSF was investigated as an experimentally accessible probe for scanning tunneling spectroscopy (STS) measurements of strongly correlated electron systems\cite{Ervasti2017c}.
The combined resolution of charge excitations in real and frequency space, permits to comprehensively elucidate the fate of single-particle edge states in the presence of interactions. At half filling, it was shown that the edge excitations of the SSHH model remain quasi-particle like for moderate on-site interactions and allow clear experimental distinction of the trivial and topological phases of the SSHH model\,\cite{Mikhail2022d}. 
As the impact of on-site interactions is strongest for the edge sites, at sufficiently large interactions bulk occupation becomes energetically favorable and the quasi-particle picture breaks down.

\section{SSHH model at quarter filling}\label{sec:qf}

In this work, we focus on the less studied quarter-filled case of the SSHH model. While the 1D Hubbard model ($\dt=0$) as well as the non-interacting SSH model ($U=0$) are both metallic at quarter filling ($n=1/4$), the combination of dimerization and strong on-site interactions leads to a situation akin to a half-filled single-band system\,\cite{Penc1994,Nishimoto2000} due to the formation of singly occupied dimers, \ie the pairs of bulk sites connected by strong bonds $t_+$ in \fig{f:u}b.
With other words, analogous to a half-filled metallic band undergoing a Mott-transition, we deal with a half-filled lower band, corresponding to quarter filling.
The associated ground state features a Mott gap and anti-ferromagnetic correlations\,\cite{Le2020}. 
    
In contrast to half-filling, at strong on-site interactions edge occupation is favored over bulk occupation in the quarter-filled system, which leads to clear signatures of quasiparticle edge excitations in the SPSF.\@ Moreover, since the edge occupation is only found for $\dt>0$, the SPSF can be used to distinguish the two dimerized configurations. 

As observed earlier, the quarter-filled Hubbard model maps to the half-filled spinless tight-binding model for $U\to \infty$\,\cite{Ovchinnikov1973}. Even away from the exact limit, this relation can be utilized to realize the effectively non-interacting spinless SSH model by means of an external magnetic Zeeman field\,\cite{Le2020}. 
Finally, the presence of a large bulk gap makes quarter filling experimentally accessible as demonstrated by its recent implementation in engineered dopant lattices\,\cite{Kiczynski2022b}. 
    
An overview of the behavior of the SSHH model at different fillings is provided in \fig{f:u}\,a which shows the addition spectrum\md a generalization of the band structure to interacting systems\md for $\dt = 0.5 $ and $L=12$. The addition energies are defined as $E_+(N)=E^N_0-E^{N-1}_0$ between different fillings $n=N/(2L)$ of $N$ electrons and $L$ lattice sites; $E^N_0$ is the $N$-particle ground state energy. 

Similar to the band gap in non-interacting systems, a gap in the addition spectrum, also referred to as {\it charge gap}, is defined as $\Delta_c(N)=E_+(N+1)- E_+(N)$; it implies an insulating $N$-particle ground state. 
To emphasize the particle-hole symmetry (PHS) of the model, the chemical potential is set to $\mu=U/2$. The color-coding in Fig.\,\fig{f:u}\,a indicates the edge occupation $\braket{n}_\text{edge}=\braket{\hat{n}_1} + \braket{\hat{n}_L}$.  
    
\begin{figure*}[t!]
\begin{center}
    \centering
    \vspace*{0.5cm}   
    \includegraphics[width=\linewidth]{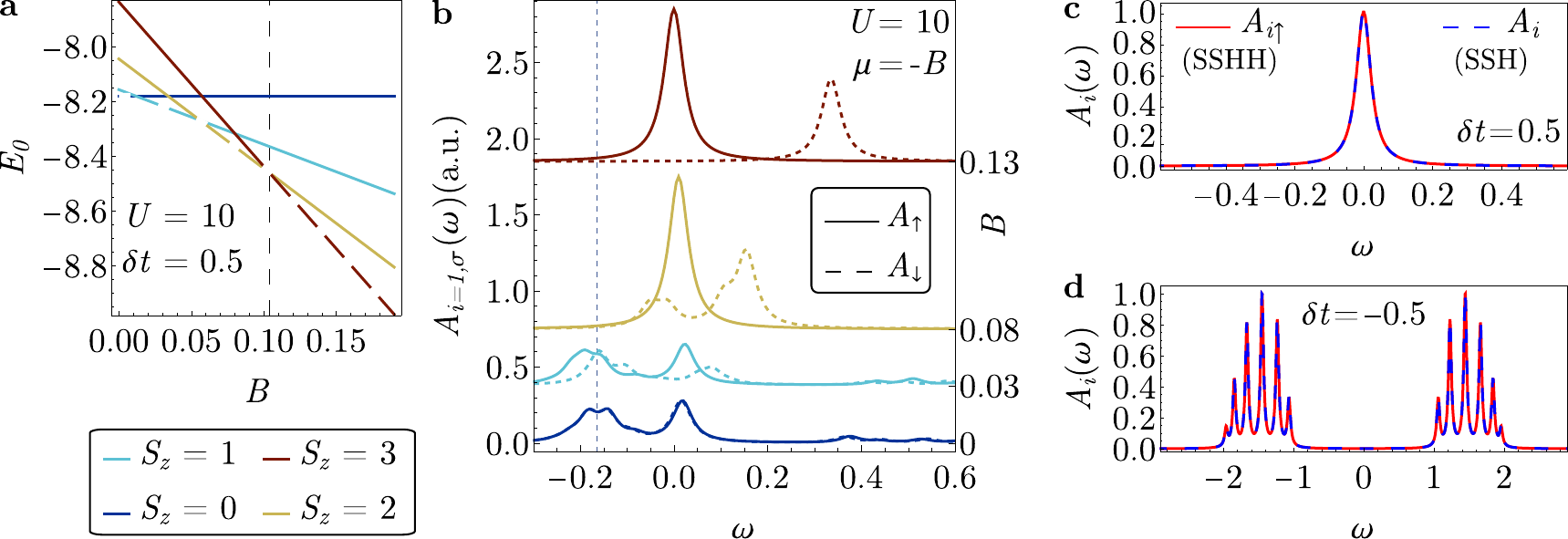}
    \caption{\textbf{Magnetization of ground state and spectral function for finite magnetic field at quarter filling.} \textbf{a} Ground state energy and magnetization in dependence of magnetic field. Solid lines correspond to different magnetization sectors while dashed line corresponds to overall ground state energy for a given value of $B$. The vertical dashed line indicates the transition to the sector with saturated magnetization.\ \textbf{b} Spectral function of an edge site at different magnetic field strengths corresponding to different ground state polarizations. Both plots are calculated for $L=12$, $U=10$ and $\delta t = 0.5$.\ \textbf{c} Spin-up component of fully polarized ($S_z=3$) edge spectral function at quarter filling (yellow) and $U=10$ compared to edge spectral function of spinless SSH model ($U=0$) at half filling (blue, dashed). The former is calculated for $\mu = -B $ to account for the constant shift in energy due to the magnetic field.\ \textbf{d} Same as \textbf{c} but in the topologically trivial phase ($\delta t = -0.5$).}\label{f:ub}
\end{center}
\end{figure*}

The three different fillings at which charge gaps appear \md half filling, quarter filling and three-quarter filling\md  are highlighted. Since the latter two are related by the particle-hole transformation $\hat{c}_{j,\sigma} \to (-1)^j\hat{c}^\dagger_{j,\sigma}$ which leaves the Hamiltonian invariant and leads to $E_+(N)\to E_+(L-N)$, in this work we only focus on quarter filling.

Starting from a metallic system at $U=0$, where the Fermi-energy of the quarter filled SSH model lies at the middle of the half-filled lower band, a Mott gap\,\cite{Nishimoto2000} develops with growing values of $U$. In the thermodynamic limit and strong coupling regime, the charge gap approaches the value of the non-interacting dimerization gap $\Delta_\delta=4\dt$\,\cite{Penc1994,Nishimoto2000}. Notably, the gap size is substantially smaller than for the half-filled case, as the lower density allows electrons to re-arrange as to minimize double occupation.     

The increase in $\braket{\hat{n}}_\text{edge}$ with $U$ at quarter filling can be intuitively understood in the fully dimerized limit ($\dt=1$) where the chain decouples into two isolated edge sites and separated two-site Hubbard systems in the bulk, which we refer to as dimers. Whether an edge or a bulk site is occupied depends on the charge gap of the associated excitations. The energy cost to add a second electron to a bulk dimer is $\Delta_{\rm dimer}= E^{N=2}_{\rm d} + E^{N=0}_{\rm d} - 2E^{N=1}_{\rm d}$ with dimer energies $E^{N=0}_{\rm d} =0$, $E^{N=1}_{\rm d}=-t_+$ and $E^{N=2}_{\rm d}=(U-\sqrt{U^2+16(t+\dt)^2})/2$. Since the edge states have zero energy $E^{N=1}_{\rm e}=0>E^{N=1}_{\rm d}$, all bulk dimers are singly occupied while the edge sites are empty when the system is one electron away from quarter filling. An added electron will occupy the edges once the dimer gap becomes positive, \ie $\Delta_{\rm dimer }> E^{N=1}_{\rm e} = 0$. The average occupation at quarter filling for an edge (red) and bulk (blue) site, respectively, are shown in \fig{f:u}c. In the fully dimerized limit (dashed lines), the edge occupation per site jumps discretely from $\braket{n_i}=0\to 1$ at $U=6$, for which $\Delta_{\rm dimer }=0$. 
Away from the fully dimerized limit at $\dt=0.5$, which corresponds to the addition spectrum in \fig{f:u}\,a, the edge and bulk states hybridize and the average edge occupation per site increases gradually as a function of $U$. 

In \fig{f:u}\,d we show the local probability density for electron addition $\Delta n_i(N) = \braket{\Psi^{N+1}_0|\hat{n}_i|\Psi^{N+1}_0}-\braket{\Psi^N_0|\hat{n}_i|\Psi^N_0}$ at quarter filling ($N=6$) in dependence of $U$. As per inversion symmetry, we only present the first half of the chain.
At $U=0$, charge excitations on the edges are strongly suppressed and are mostly centered on bulk dimers with the exception of the central dimer of sites 6 and 7. 
As shown in the inset, with increasing $U$ the probability of edge occupation supersedes the probability of bulk occupation. This shift from bulk to edge occupation is in contrast to the half filled case where the bulk occupation increases with $U$, with the strongest increase found on the central dimer of the chain\,\cite{Mikhail2022d}. 

The spectral function, obtained by summation of the local spectral function $A_i(\omega)$ over the chain, 
\begin{align}
    A(\omega) = \sum_i A_i(\omega),         
\end{align}
is shown in \fig{f:u2}a and \fig{f:u2}b at quarter filling. At $U=0$, it is identical to the density of states (DOS).
    
In this non-interacting limit, the excitation energies correspond to the single-particle spectrum of the SSH model shown in the inset of \fig{f:u2}c. The dimerization opens a charge gap at $n=1/2$, which we call the dimerization gap $\Delta_\delta$,  while at quarter filling the system is metallic with a half filled lower band, indicated by the addition $E_+$ and removal $E_-$ energies. 
    
For $U>0$, a charge gap opens in the lower (half-filled) band and the upper half of the lower band is pushed towards the upper band which leads to the shrinking of the dimerization gap $\Delta_\delta$. The upper SSH band itself splits into a band at $\omega \approx U$ (not shown here) which involves double occupations and a low-energy band which resembles the initial conduction band at $U=0$. 
    
Interestingly, the addition and removal energies, which are the excitation energies between ground states, behave differently for positive and negative $\dt$. 
At $\dt>0$ (\fig{f:u2}a), $E_\pm$ both move into the charge gap of the lower band. These correspond to edge excitations and are the reason for the increase in $\braket{\hat{n}}_\text{edge}$ in \fig{f:u}a. 
The spectral function for $\dt<0$ is shown in \fig{f:u2}b. Here, the ground state excitations $E_\pm$ are separated by the newly formed charge gap. In this case the chain consists of $L/2$ dimers, each occupied by approximately one electron. Electron removal corresponds to an excitation energy of $E_0^N-E_0^{N-1}$ which, for large dimerizations, approaches $\approx E_d(N=1) = -(t+\dt)$. In contrast, charge addition involves doubly occupied dimers with a $U$-dependent energy. 
    
The low-energy excitations of the $\dt<0$ case resemble the spectral function of a half-filled (non-dimerized) Hubbard model with renormalized couplings\,\cite{Nishimoto2000}. For large and negative $\dt\ll 0$, we can approximate the dimers as single sites of a regular Hubbard chain of length $L/2$ with effective hopping $(t-\dt)/2$ and on-site interaction $U\approx \Delta^{\rm dimer}_c = E_d(N=2)+E_d(N=0)-2E_d(N=1)$, where $\Delta^{\rm dimer}_c$ denotes the charge gap of an isolated dimer. The corresponding spectral function is shown in \fig{f:u2}c in blue compared to the quarter filled SHHH model (red). The center of the charge gaps is aligned to ease comparison. For both models the SPSF features two Hubbard bands separated by charge gaps of similar sizes. Due to the smallness of the gap between the  ``upper Hubbard band'' around quarter filling and the residual SSH band, hybridization leads to asymmetric peaks around the charge gap, which is not present in the half-filled Hubbard model.
The formation of an effectively half-filled system at low energies can serve as a benchmark signature for experiments which attempt to simulate the dimerized Hubbard model. 
    
In the following, we will study the effect of a magnetic field $B$ and nearest neighbor interactions $V$ on the quarter-filled SSHH model, with focus on the strongly correlated limit $U/t=10$, which lies well within the range of the experimentally found values of $U/t\approx 4\nd 16$, with average interactions of $U/t\approx 6.6$ and $U/t\approx 9.1$ for positive dimerization $\dt \approx 0.58$ and negative dimerization $\dt \approx -0.35$\,\cite{Kiczynski2022bc}.

\begin{figure*}[t!] 
    \begin{center}
        \centering
        \vspace*{0.5cm}   
        \includegraphics[width=\linewidth]{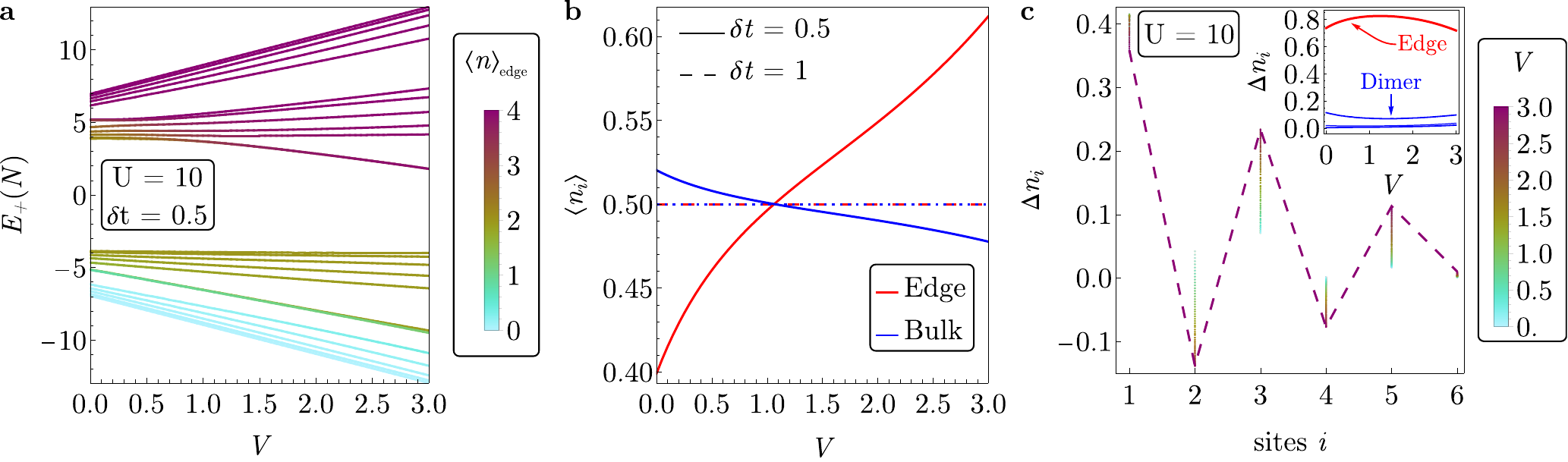}
        \caption{\textbf{Extended SSHH model of an open chain for strong on-site interactions $U=10$.} \textbf{a} Addition spectrum in dependence of $V$. Each line corresponds to a different particle number $N$ and the color indicates the edge occupation $\braket{n}_\text{edge}=\braket{\hat{n}_1} + \braket{\hat{n}_L}$.\ \textbf{b} Average occupation of an edge (red) and a bulk (blue) site, respectively. Solid lines correspond to $\dt=0.5$ while dashed lines are calculated for $\dt=1$.\ \textbf{c} Local charge density profile $\Delta n_i = \braket{\hat{n}_i(N=7)}-\braket{\hat{n}_i(N=6)}$ at quarter filling. Dashed line highlights the charge distribution for the largest $V$ calculated (here $V=3$). Inset: Average charge distribution of edge sites (red) and bulk dimers (blue) in dependence of $V$.}\label{f:uv}
    \end{center}
\end{figure*}

\section{SSHH model in a magnetic field\label{sec:ub}}
In Ref.\,\cite{Le2020c}, the SSHH model at quarter filling is predicted to map to the spinless half-filled SSH model in the presence of a sufficiently strong magnetic field, for which the ground state magnetization is saturated, \ie the system is populated by a single spin-species. Since in this case double occupation is prevented by the Pauli exclusion principle, the eigenvalues of $\hat{H}_U$ in \eq{hu} vanish and the system is effectively described by the spinless version of the non-interacting tight-binding Hamiltonian \eq{hssh}. 
The two-fold degenerate ground state as well as the drop in entanglement entropy between edge and bulk sites for $\dt>0$ found by the authors of Ref.\,\cite{Le2020c} is consistent with an emergent SSH model. While the presented results are promising, further evidence may be desirable to exclude other explanations for these findings. 

To substantiate this idea and to provide an experimentally accessible observable, we show calculations of the SPSF in the presence of an external magnetic field generated by the Zeeman term \eq{hb}. First, we assume $\vv=0$ and leave the $\vv >0$ discussion to \sec{sec:ubv}. 
Note that the here presented topological signatures in the SPSF are readily observable as many STM experiments are equipped with magnets of up to several Tesla. 
    
As shown in \fig{f:ub}a, the magnetization of the ground state increases with $B$. The maximum magnetization is given by $S^\text{max}_z=N/2 \equiv L/4$, where the last equivalence holds for quarter filling. We denote the magnetic field for which the spin-polarization is saturated as $B_{\rm sat}$. In general, $B_{\rm sat}$ depends on the choice of system parameters. For the $L=12$ chain with $U=10$ and $\dt=0.5$ we find $B_{\rm sat}/t \approx 0.1$. 

Since the magnetic field couples to the electron spin, we show its effect on the spin up (solid) and spin down (dashed) edge SPSF separately in \fig{f:ub}b. Each curve is calculated for magnetic field strengths (indicated on the right axis) corresponding to different ground state magnetizations $S_z$. The latter are indicated by the color of the curves. Each curve is vertically shifted by hand to increase readability as well as horizontally by $\mu=-B$ to compensate for the constant energy shift in the fully polarized case (red) where $N_\up = N$.

At $B=0$ and $S_z=0$ (dark blue), the strongly correlated edge SPSF features a double peak structure at low energies. The left peak corresponds to ground state transitions, \ie $\ket{\Psi^N_0} \to \ket{\Psi^{N\pm1}_0}$. The right peak is caused by charge excitations involving the addition of an electron into higher excited states, \ie $\ket{\Psi^N_0} \to \ket{\Psi^{N+1}_n}$ with $n>0$. 

At $B=0.03$, the ground state magnetization increases $S_z = 0 \to 1$ and the spin up $A_{i,\up}$ and spin down $A_{i,\dw}$ components split. The addition peaks of the latter move to higher excitation energies as the addition of spin-down electrons is energetically penalized by the magnetic field. 

For $B=0.08$ the spectral weight of $A_{i,\up}$ is mostly localized in the low-energy peak corresponding to ground state transitions while the high-energy peak is strongly suppressed. The separation between the spin-components becomes more pronounced with $A_{i,\dw}$ showing substantial transfer of spectral weight from the low-energy peak to high-energy excitations, including small emerging peaks around $\omega\approx1$.

Finally, at $B=0.13>B_{\rm sat}$, the ground state is fully polarized, \ie $S_z=3$, with all spins aligned with the magnetic field. Ground state transitions in $A_{i,\dw}$ as well as high-energy excitations in $A_{i,\up}$ are now fully suppressed. In particular, the ground state transitions of the spin up components are centered around $\omega=0$ (for $\mu=-B$), which matches the edge SPSF in the topological phase of the non-interacting SSH model.

The comparison between the spin up polarized SPSF of the interacting SSHH model at quarter filling (yellow) and the SPSF of the spinless non-interacting SSH model at half filling (blue, dashed) is depicted in \fig{f:ub}c\nd d for the topologically non-trivial ($\dt=0.5$) and trivial ($\dt=-0.5$) phases, respectively. In both cases the interacting and non-interacting spectral functions show perfect agreement, which verifies the proposed\,\cite{Le2020c} magnetic-field induced transition between the SSHH model at quarter filling and the spinless SSH model at half-filling.

\begin{figure*}[t!]
    \begin{center}
        \centering
        \vspace*{0.5cm}   
        \includegraphics[width=\linewidth]{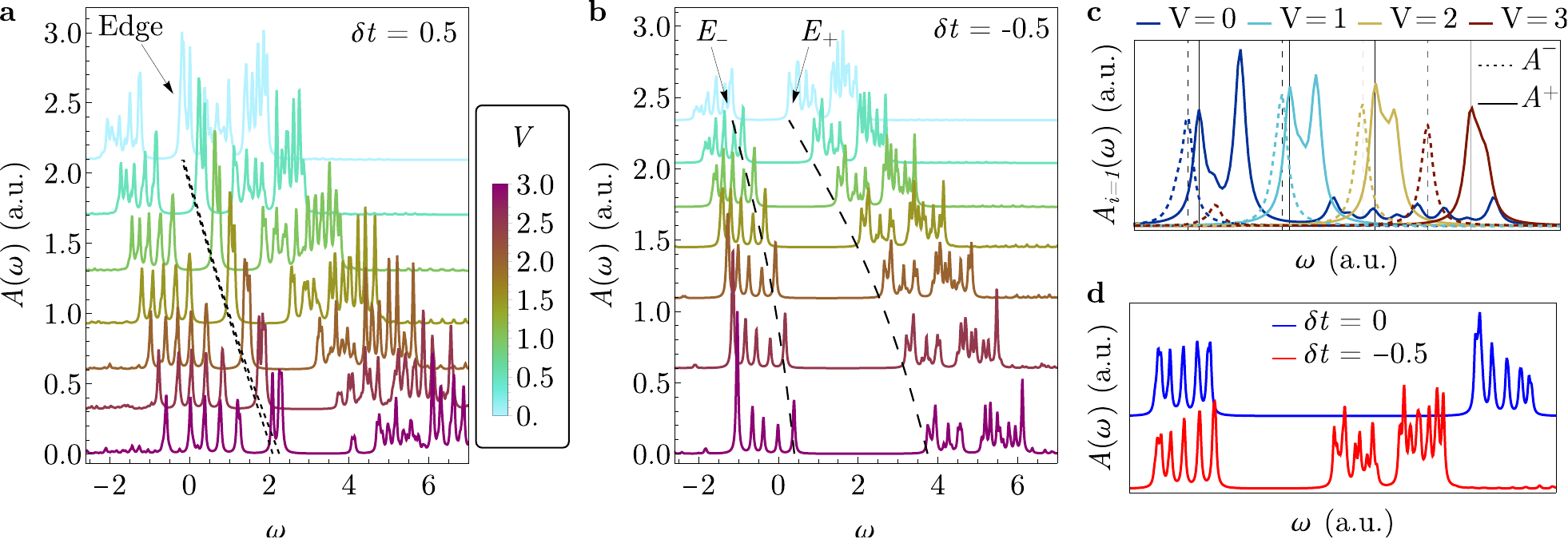}
        \caption{\textbf{Spectral functions of the extended SSHH model.} \textbf{a} SPSF $A(\omega)$ for $\dt = 0.5$ summed over all sites. The edge excitations within the gap are indicated by arrow for $V=0$ (cyan). $V$ increases in steps of 0.5 for successive curves, with $V_{\rm \max} = 3$. The dashed curves indicate addition $E_{+} (N)$ and removal $E_{-} (N)$ energies.\ \textbf{b} Same as \textbf{a}, but for $\dt=-0.5$.\ \textbf{c} Addition (solid) and removal (dashed) SPSF of edge site for $\dt=0.5$ and $V=0,1,2,3$.\ \textbf{d} SPSF of half-filled extended Hubbard model (blue) and quarter-filled extended SSHH model (red). The former is calculated for $t=0.25$, $U=5.5$ and $V=0.25$ (see main text), the latter is calculated for $t=1$, $U=10$ and $V=1$. Lower bands of both curves are aligned to improve visibility.}\label{f:uv2}
    \end{center}
\end{figure*}

\section{Extended SSHH model\label{sec:uv}}

In this section we discuss the effect of nearest-neighbor interactions on the SSHH model at quarter filling. The additional nearest-neighbor repulsion gives rise to the attribute ``extended''.
In general, the competition of $V$ with $U$ and $\dt$ might lead to novel or additional phases such as charge-density wave (CDW) instabilities; longer-range Coulomb interactions are inevitably present in analogue quantum simulators such as dopant lattices on silicon, which recently were found to be $V/t\approx 0.5\nd1.8$, with average values of $V/t\approx 0.7 $ and $V/t\approx 1 $ for positive and negative dimerization, respectively \,\cite{Kiczynski2022bc}. 

In the limit $U\to\infty$, the extended 1D Hubbard model ($\dt=0$) at quarter filling undergoes a metal-insulator transition at $V=2t$\,\cite{Ovchinnikov1973}, where the insulating phase at $V>2t$ features a charge density wave (CDW)\cite{Penc1994a}. The transition persists even away from the exact limit for finite $U$ and $V$\,\cite{Mila1993}, which is in contrast to the half-filled case, where the system is insulating for any $U>0$. Apart from CDW, the inclusion of nearest-neighbor interactions can lead to a variety of phenomena such as spin density waves (SDW), superconducting fluctuations and phase separation\,\cite{Lin1995,Clay1999}.
    
While the metal-insulator transition is absent in the extended SSHH model, since the combination of dimerization and on-site interactions generates a charge gap even for $V=0$, finite $V>0$ lead to a substantially stronger increase of the charge gap compared to on-site interactions. 

For large dimerization $t_{+}\gg t_{-}$, the spin sector is gapless as long as $U>0$, since in this limit the model maps to the half filled extended Hubbard model with $L/2$ sites and renormalized couplings\,\cite{Nishimoto2000}. The absence of a spin gap renders the extended SSHH model at quarter filling topologically trivial. 
Moreover, the tendency of $V$ to form charge density waves, together with inversion symmetry, leads to a growing edge occupation.

The addition spectrum for the experimentally relevant case of strong on-site interactions, here $U=10$, and weak to moderate nearest-neighbor interactions is presented in \fig{f:uv}a. The chemical potential is set to $\mu=U/2+2V$ to compensate for $U$ and $V$ dependent energy shifts that are linear in $N$. The PHS is broken for $V>0$ as clearly visible from the asymmetrically shrinking charge gap at half filling. 

At quarter filling, on the other hand, $V$ substantially drives the growth of the charge gap and edge occupation; the latter is depicted (red) in \fig{f:uv}b together with the average bulk occupation (blue) for $\dt=0.5$ (solid) and $\dt=1$ (dashed). In the fully dimerized limit we find $\braket{n}_{\rm edge} = 0.5$ since only one edge site is occupied as to avoid occupation of neighboring sites. Away from this limit, the edge states hybridize resulting in an increased edge occupation that exceeds the average occupation per bulk site at around $V \approx 1$. 

The effect of $V$ on the charge density distribution upon electron addition is shown in \fig{f:uv}c for $U=10$ and $\dt = 0.5$. The sublattice polarization, due to the dimerization $\dt$, strongly increases with $V$ and is visualized as a dashed line for $V=3$. The rearrangement of charge distribution occurs mostly within the bulk dimers, while little charge density is exchanged between dimers as shown in the inset of \fig{f:uv}c. For $V\leq3$, both densities of the edges and of individual bulk dimers are weakly affected. 
The edge occupation increases by up to $\approx 12\%$ at $V\approx 1.25$ followed by a decrease of $\approx 2.5 \%$ at $V=3$ compared to the value without nearest-neighbor interactions. 

The SPSF of the extended SSHH model is shown in \fig{f:uv2}a and b for positive and negative dimerization $\dt=\pm0.5$, respectively. Both plots are calculated in the strong coupling regime $U=10$. Successive curves correspond to increasing values of $V$ in steps of 0.5, starting from $V=0$ at the top. The dashed curves show the addition $E_+$ and removal $E_-$ energies, which correspond for $\dt=0.5$ to edge excitations within the charge gap of the lower band, while at negative dimerization the ground state excitations remain in separated bands. In both cases, the charge gap between the bands grows strongly with $V$. 
\begin{figure*}[t!]
    \begin{center}
        \centering
        \vspace*{0.5cm}   
        \includegraphics[width=\linewidth]{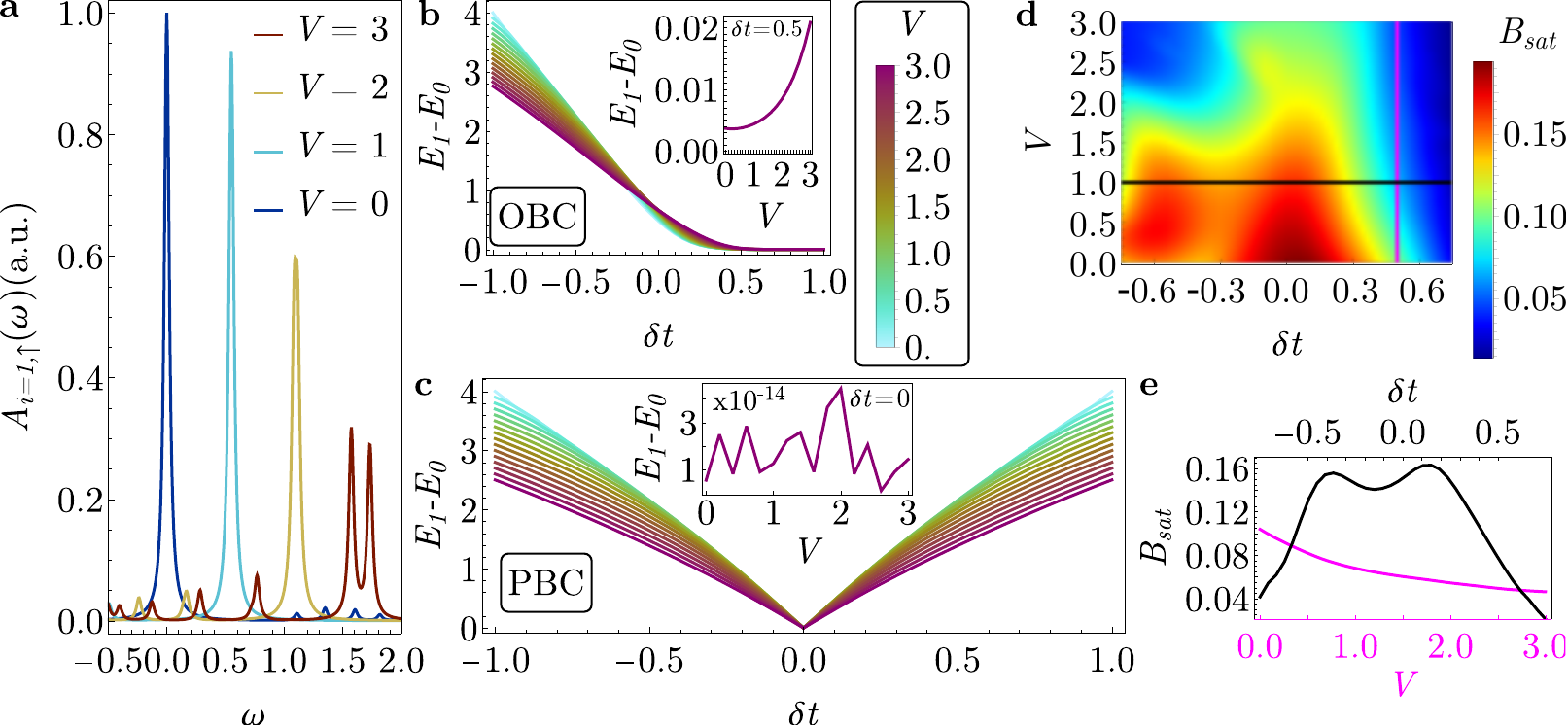}
        \caption{\textbf{Effect of nearest-neighbor interactions $V$ on the fully polarized state at $B>B_{\rm sat}$}. All observables are calculated at quarter filling for $U=10$, $B=0.13$ and $S_z=3$ if not stated otherwise.\ \textbf{a} Edge spectral function for various $V$. The chemical potential is set to compensate the overall shift of the magnetic field.\ \textbf{b} Eigenenergy gap in dependence of dimerization $\delta t$ for open boundary conditions (OBC). Different colours correspond to varying $V$. The inset shows the gap size for $\dt=0.5$ in dependence of $V$.\ \textbf{c} Same as \textbf{b} but for periodic boundary conditions (PBC). The inset depicts the eigenenergy gap at $\dt=0$, \ie the gap closing point, in dependence of $V$.\ \textbf{d} Saturated magnetic field $B_{\rm sat}$ at $U=10$ in dependence of $\dt$ and $V$.\ \textbf{e} $B_{\rm sat}$ along cuts in \textbf{d} for constant $\dt=0.5$ (pink) and constant $V=1$ (black), respectively.}\label{f:uvb}
    \end{center}
\end{figure*}

The dominant double-peak structure of the edge excitations is depicted in more detail in \fig{f:uv2}c, where the addition (solid) $A^{+}_1(\omega)$ and removal (dashed) $A^{-}_1(\omega)$ components of the edge SPSF are shown for several values of $V$. The vertical lines correspond to the addition (solid) $E_+$ and removal (dashed) $E_-$ excitation energies between the respective ground states.
The left peak corresponds to ground state transitions for both electron addition and removal, while the right peak is due to electron addition into higher excited states (here $\{E^{N+1}_n\}_{n=2,\dots,5}$). 

For weak to moderate interaction strengths $V$, spectral weight transfer occurs from the right to the left peak as the addition energy $E_+$ approaches the removal energy $E_-$, leading to a growing amplitude of the ground state excitations. At $V\approx 1$, the charge gap on the edge reaches its minimum and the left peak becomes maximal. For larger $V$ the addition and removal parts move apart, separated by a growing charge gap, which potentially indicates a phase transition in the thermodynamic limit.

As discussed in \sec{sec:qf}, for large dimerization and strong on-site interactions, the lowest to bands of the quarter-filled system resemble Hubbard bands at half filling with renormalized couplings. In the presence of nearest-neighbor interactions, the quarter-filled extended SSHH model\md characterized by the parameters $L,\, t, \, \dt, \, U$ and $V$\md is related to the half-filled Hubbard model with $L'=L/2, \, t'=(t-\dt)/2, \, U'=(U+V)/2$ and $V'=V/4$.\,\cite{Nishimoto2000}. The SPSF of each model is shown in \fig{f:uv2}d for $U=10$ and $V=1$. Both curves are aligned w.r.t.\ the lower band to ease the visual comparison. We find a striking similarity of the lower bands of both models, both in peak structure and bandwidth. The upper bands are less similar, which is likely due to the proximity to a third high-energy band in the quarter-filled case. Furthermore, the gap sizes of both models differ which is expected due to finite-size corrections\,\cite{Nishimoto2000}. Nevertheless, the Hubbard-type phenomenology is present and provides a clearly visible benchmark for experimental realizations of the extended SSHH model at quarter filling.

\section{Extended SSHH model in a magnetic field\label{sec:ubv}}

In this section we consider the effect of $V$ on the spin-polarized SPSF of the fully magnetized ground state at quarter filling discussed in \sec{sec:ub}. 

In \fig{f:uvb}\,a we show the spin up component of the edge SPSF for different values of $V$. The chemical potential is set to $\mu=-B$ to compensate the overall shift due to the magnetic field. While the peak amplitude decreases with growing $V$, the quasi-particle nature of the SSH-like peak remains visible for $V\leq2$. Only for sufficiently large $V$ (here $V=3$) the addition and removal peak are shifted to higher and lower energies, respectively. Furthermore, small side-peaks of charge removal excitations appear at $\omega \approx - 1$. 

Despite the deviation of the SPSF from the zero-energy excitation of the non-interacting SSH model, analysis of the many-body eigenenergies indicates the persistence of the topological phase. In \fig{f:uvb}\,b we show the many-body gap $\Delta E= E_1-E_0$ between ground state and first excited state of an open chain. In the thermodynamic limit the eigenenergy gap of an open SSH chain is known to close when the dimerization $\dt$ changes sign due to the emergence of degenerate zero-energy edge states. In a finite system, the critical dimerization value is shifted away from $\dt=0$, nevertheless the transition appears to be stable for $V>0$. 
The effect of $V$ is to shrink the gap in the topologically trivial phase and to further increase the transition value. As shown in the inset, for large $V$, the size of the many-body gap is comparable to the peak-broadening $\eta=0.025$ and the the two lowest eigenstates can no longer be considered degenerate at $\dt=0.5$, explains the energy splitting of the red curve in \fig{f:uvb}\,a.  

Another sign for the persisting topological phase transition is evidenced by the closing and re-opening of the many-body gap at PBC as shown in \fig{f:uvb}\,c. In accordance with the non-interacting SSH model, the gap closes and re-opens at the critical value $\dt=0$. Remarkably, we find $\Delta E(\dt=0) = 0$ independent of the interaction strength $V$ (\cf inset \fig{f:uvb}\,c), which is in contrast to the half filled SSHH model\,\cite{Mikhail2022d} where, due to finite-size effects, the gap closing point of the non-interacting model becomes a local but finite minimum of the many-body gap in the interacting system. 

The dependence of the critical magnetic field $B_{\rm sat}\equiv B(S_z^\text{max})$, \ie the minimum field strength for which the ground state polarization is saturated, is shown in \fig{f:uvb}\,d in dependence of dimerization and nearest-neighbor interactions. The effect of $V$ is to further lower the required saturation field by a factor of $\sim 2.25$ down to $B_{\rm sat}\approx 0.046$. Assuming an average hopping amplitude of $t\approx 1\rm meV$ and strong on-site interactions of $U/t=10$, the saturated magnetic field drops from $B\approx 1.8 \rm T$ at $V/t=0 $ to $B\approx 0.8 \rm T$ at $V/t=3$. In the experimentally relevant range of $V/t\approx1$, the magnetic field lies at $B\approx 1.2 \rm T $, which is readily within reach for most magnets of current STMs.  
We further find that, for moderate $V$, the lowest values of $B_{\rm sat}$ are reached at positive dimerization when the system becomes topologically non-trivial.

\section{Conclusion\label{sec:conclusion}}

Motivated by its recent realization in dopant lattice experiments we study the SSH model in the presence of on-site and nearest-neighbor interactions which both are inherent to these analogue quantum simulators. We further analyze the effect of an external magnetic Zeeman field which can be easily realized in artificial lattices. 

To this end we study several observables, most importantly, the experimentally relevant single-particle spectral function.
We focus on quarter filling which is characterized by a large charge gap generated by both types of interactions and which can experimentally be readily realized. For strong dimerization, we find that the SPSF resembles the low-energy excitations of the half-filled Hubbard model with renormalized couplings. 

In the presence of sufficiently strong magnetic field for which the ground state magnetization is saturated, an effectively non-interacting spin-polarized SSH model emerges. For open chains, we present the appearance of the characteristic edge excitations of the SSH model in the spin-polarized SPSF.\@ We find perfect agreement with the SPSF of the SSH model for both $\dt\lessgtr0$. 

In the absence of a magnetic field we find that nearest-neighbor interactions lead to CDW like charge distribution in the bulk. Interestingly, for experimentally realistic values, we find nearest-neighbor interactions to have a stabilizing effect on charge excitations on the edge as evidenced by increased charge densities and narrowing quasi-particle peaks in the SPSF.\@ For large $V$, the charge gap on the edge widens, indicated by separating in-gap peaks.

Finally, we show that the SSH-like edge excitations at strong magnetic fields are robust against $V$ within the experimentally expected range. The stability of the topological phase transition is confirmed by the analysis of both OBC and PBC excitation gaps. Moreover, we find that the required magnetic field to reach full polarization and to realize the effective SSH model is lowered by the presence of $V$, making the proposed phase experimentally more accessible.

\begin{acknowledgements}
We acknowledge discussions with S.\ Rogge, G.\ Buchs, B.\ Voisin and D.\ Didier St Medar.
S.R.\ acknowledges support from the Australian Research Council through Grant No.\ DP200101118 and DP240100168.
This research was undertaken using resources from the National Computational Infrastructure (NCI Australia), an NCRIS enabled capability supported by the Australian Government.
\end{acknowledgements}

\bibliographystyle{prsty.bst}
\bibliography{references/references}

\end{document}